\documentclass[reprint,superscriptaddress,
amsmath,amssymb,aps,pra,floatfix,
]{revtex4-2}
\usepackage{graphicx}
\usepackage{dcolumn}
\usepackage{bm}
\usepackage{amssymb}
\usepackage{lipsum}
\usepackage{braket}
\usepackage{bbm}
\usepackage{mdframed,xcolor}
\usepackage{xcolor}
\usepackage{tcolorbox}
\definecolor{myshade}{RGB}{255,235,235}
\usepackage{colortbl}
\usepackage{tikz} 
\usepackage[colorlinks=true,linkcolor=blue, citecolor=red, urlcolor=blue]{hyperref}
\usepackage{url}
\begin{document}
\preprint{APS/123-QED}
\preprint{APS/123-QED}
\title{Effective SU(2) gadget: holonomic walk on higher-order Poincar\'{e} sphere}

\author{Mohammad Umar}
\email{aliphysics110@gmail.com}
\affiliation{Optics and Photonics Centre, Indian Institute of Technology Delhi, New Delhi 110016, INDIA}
\author{Sarvesh Bansal}
\email{bansalsarvesh.s@gmail.com
}
\affiliation{Department of Physics, Inidian Institute of Technology Delhi, New Delhi 110016, INDIA,\\
Current address: Dipartimento di Fisica, Universit\`{a} di Napoli Federico II, Complesso Universitario di Monte Sant’Angelo, Napoli 80126, ITALY
}
\author{Paramasivam Senthilkumaran$^{1,}$} 
\email{psenthilk@yahoo.com}
\noaffiliation
\begin{abstract}
In a manner commensurate to the SU(2) gadget for the Poincar\'{e} sphere, which involves a combination of two quarter-wave plates and one half-wave plate regardless of their sequential order, an analogous construct for the higher-order Poincaré sphere had long remained elusive. To address this, we recently demonstrated, by modifying Euler-angle parameterization, that an optical gadget consisting of two quarter-wave $q$-plates and one half-wave $q$-plate, each endowed with the same topological charge $q$, operates as a viable SU(2) gadget for the higher-order Poincar\'{e} sphere, contingent upon the fulfillment of the holonomy condition. This work presents the controlled navigation on the higher-order Poincaré sphere through the proposed architecture, formulated under the concept of an effective waveplate and thus can be referred to as an effective SU(2) gadget. The notion of an effective waveplate refers to a coaxial arrangement of multiple waveplates that, under specific constraints, can act as a single waveplate. In this gadget, the relative alignment of the offset angles of the constituent $q$-plates emerges as the decisive parameter governing systematic navigation on the higher-order Poincaré sphere. This study bear direct relevance to the deterministic control and engineering of structured light, encompassing polarization singularities, vector vortex beams and topological optical fields. Moreover, these results holds potential applications in the contemporary research frontiers in singular optics, spin-orbit photonics and quantum communication.
\end{abstract}
\maketitle
\section{Introduction}
\begin{figure*}[t]
    \centering
    \includegraphics[width=0.985\linewidth]{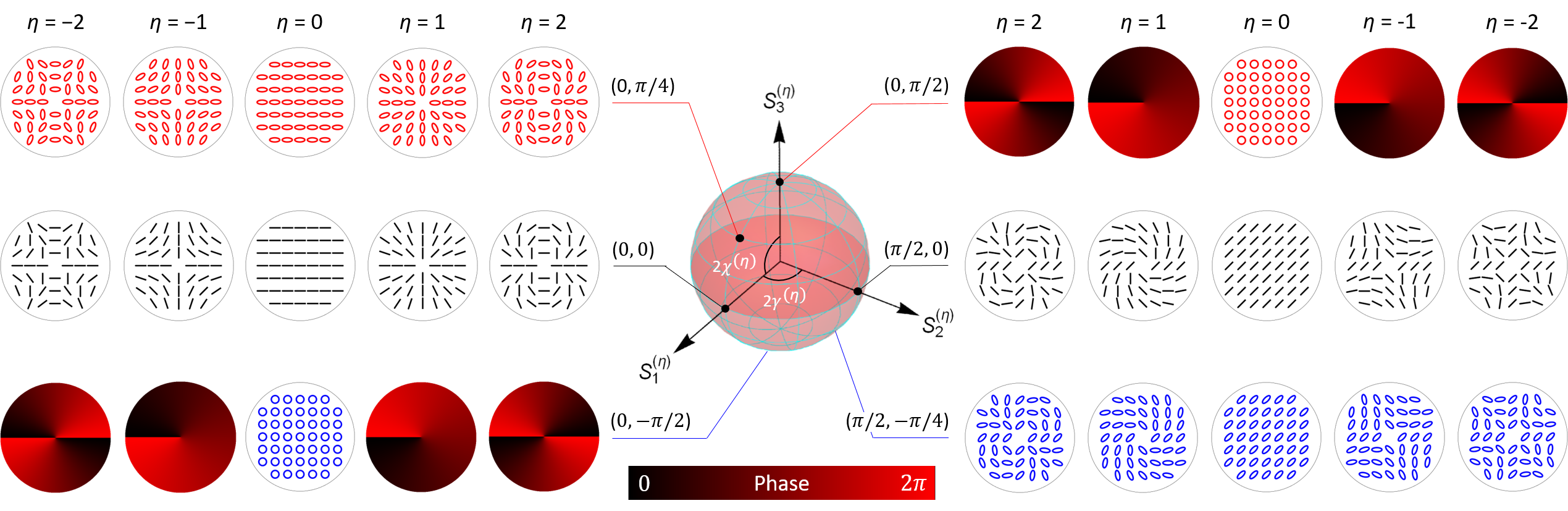}
    \caption{(Color online). The polarization distribution on the higher-order Poincaré spheres for $\eta = -2$, $-1$, $0$, $+1$, and $+2$ is displayed. Here, $2\gamma^{(\eta)}$ and $2\chi^{(\eta)}$ represent the longitude and latitude coordinates, respectively. The red and blue color coding indicate right circular polarization (RCP) and left circular polarization (LCP), respectively. Additionally, the vortex phase embedded with the RCP and LCP for the respective values of $\eta$ (except for $\eta = 0$) is also provided.}
    \label{sphere_01}
\end{figure*}

The polarization states of a plane electromagnetic wave can be mapped bijectively onto the surface of the unit $\textbf{S}^{2}$ sphere, providing a geometric representation that encapsulates a fundamental physical and mathematical insight into light polarization. This foundational insight, attributed to Henri Poincar\'{e}, led to the development of the Poincar\'{e} sphere (PS) \cite{poincare1954theorie}, which provides a powerful geometric framework for visualizing the state of polarization (SOP) of light. On the PS, each polarization state is mapped to a point on the surface of the sphere via the Stokes parameters, which serve as Cartesian coordinates (Fig. \ref{sphere_01}) \cite{stokes1851composition}. Every point corresponds to a spatially homogeneous SOP and transformations between such states are governed by the elements of the SU(2) group, a three-parameter Lie group. These SU(2) transformations are geometrically equivalent to rotations on the PS described by elements of the SO(3) group, owing to the two-to-one homomorphism between SU(2) and SO(3) groups. In optical experiment settings, birefringent optical elements such as retarders implement these SU(2) transformations and serve as experimental realizations of polarization evolution on the Poincar\`{e} Sphere.\\
\indent
Waveplates, such as quarter-wave plates (QWPs) and half-wave plates (HWPs), implement SU(2) transformations that correspond to rotations of $\pi/2$ and $\pi$ (equal to the retardance of the waveplate), respectively, on the PS and the axis of rotation is decided by the orientation of the fast axis \cite{kumar2011polarization}. Although individual waveplates cannot realize all transformations on the PS, the implementation of arbitrary rotations on the sphere necessitates a more general SU(2) element. Motivated by the need for a universal optical device capable of realizing all polarization transformations on the PS, R. Simon and collaborators published a series of three seminal papers that systematically addressed the construction of such an optical gadget. In their initial work \cite{simon1989hamilton}, it was demonstrated that a sequence of six waveplates, two QWPs and four HWPs, was sufficient to implement any arbitrary SU(2) transformation. This result was later refined \cite{simon1989universal}, where the required number of elements was reduced to four, two HWPs and two QWPs. In their final work \cite{simon1990minimal}, the authors introduced a minimal configuration consisting of only three waveplates, two QWPs and one HWP, arranged in any sequence (QHQ, HQQ and QQH). Here Q and H represent the QWP and HWP, respectively. This minimal optical setup was shown to realize a complete SU(2) tour on the PS and has since become widely recognized as a \textit{universal} SU(2) gadget for polarization optics.\\
\indent 
The PS provides a powerful geometric representation of polarization states, however, it proves inadequate for representing higher-order solutions of Maxwell’s equations that give rise to spatially inhomogeneous beams. Unlike homogeneous polarization states, which map to single points, these structured beams are represented by extended/connected regions on the PS rather than by a single point. This gap is remedied by the formulation of the higher-order Poincaré sphere (HOPS) \cite{milione2011higher}, where spatially inhomogeneous beams are represented as points on its surface. The HOPS, like the standard PS, is also topologically an $\mathbf{S}^{2}$ sphere, where the higher-order Stokes parameters serve as Cartesian coordinates. In the standard PS, the basis states are defined solely by spin angular momentum (SAM), whereas the HOPS extends this framework by incorporating both SAM and orbital angular momentum (OAM). Consequently, the HOPS allows the representation of polarized singular beams with spatially varying polarization distributions while preserving uniform ellipticity across the transverse cross-section.\\
\indent
HOPS beams arise from the superposition of right- and left-circularly polarized vortex beams carrying equal and opposite topological charges. The equator of the HOPS contains cylindrical vector beams, including canonical cases such as radially and azimuthally polarized beams. The poles correspond to right- and left-circularly polarized vortex beams of equal and opposite topological charges. The remaining points on the sphere represent spatially inhomogeneous polarization beams exhibiting uniform ellipticity across the transverse profile. The northern and southern hemispheres distinguish right-handed (red) and left-handed polarization. All these descriptions of the HOPS beams can be visualized in Fig. \ref{sphere_01}, where the polarization distributions corresponding to HOPS of orders $-2$, $-1$, $0$, $+1$ and $2$ are shown. Each order of the HOPS is associated with a distinct polarization topology, which is quantitatively characterized by the Poincar\'{e}–Hopf (PH) index \cite{freund2002polarization, ruchi2020phase, senthilkumaran2024singularities}, defined as $\eta = \frac{1}{2\pi} \oint \nabla \gamma \cdot d\mathbf{l}$, where $\gamma$ represents the
azimuth/orientation of the polarization ellipse/linearly polarized light. The line integral is taken over a closed contour enclosing a polarization singularity \cite{ruchi2020phase, senthilkumaran2024singularities}, where $\gamma$ becomes undefined. The sign of the PH index can be either positive or negative, and it indicates the handedness of the azimuthal rotation of the SOP in the vicinity of the singularity.\\
\indent
Despite the well-celebrated geometry of the HOPS and numerous studies related to it \cite{chen2014generation, naidoo2016controlled, xu2021polarization, yao2022generation, ji2023controlled, yao2023quantitative, bansal2023stokes, liu2024generation, fickler2024higher, pal2024tight, bougouffa2025optical, yang2025metasurface, he2025higher}, the realization of a full SU(2) evolution over its surface had remained elusive until our recent work \cite{umar2025mathematics, umar2025su2gadget}. We have also attempted to address this gap by theoretically engineering a \textit{general} $q$-plate \cite{umar20252} and an optical gadget where two HWPs are sandwiched between the quarter-wave $q$-plates \cite{bansal2025gadget}. By a general $q$-plate, we refer to a device with a tunable retardance spanning $0$ to $2\pi$ and a tunable offset angle ranging from $0$ to $\pi/2$. Although fabricating a plate featuring this type of tunability is currently challenging, it could become achievable using emerging metasurface and liquid-crystal approaches. The optical gadget reported in \cite{bansal2025gadget}, being a combination of homogeneous and inhomogeneous waveplates, enables only discrete hopping on the HOPS rather than a continuous polarization evolution. \\
\indent
Our latest work \cite{umar2025mathematics} presents the mathematical formalism of an effective waveplate, representing the first step toward identifying the SU(2) gadget for the HOPS. An effective waveplate refers to a coaxial arrangement of multiple waveplates that, under specific constraints, behaves as a single equivalent waveplate. Within this framework, we investigated a triadic configuration of $q$-plates, each being either a quarter-wave $q$-plate ($q^{Q}$-plate), a half-wave $q$-plate ($q^{H}$-plate) or a combination thereof. A salient result is that, under the concept of effective waveplate, any permutation of two $q^{Q}$-plates and one $q^{H}$-plate is operationally equivalent to a single effective $q$-plate endowed with a continuously tunable retardance over the full interval $[0,2\pi]$, with the tunability governed by the relative offset angles of the constituent elements. It has been shown in \cite{umar20252, umar2025holonomically, yao2023quantitative} that, under the holonomy condition, the retardance of a $q$-plate corresponds to the rotation angle on the higher-order Poincaré sphere (HOPS), while the offset angle determines the rotation axis. Consequently, with this tunable retardance, a complete $2\pi$ rotation can be achieved on the HOPS. Motivated by these findings, we subsequently demonstrated for the first time the realization of the SU(2) gadget for the HOPS \cite{umar2025su2gadget}. It has been demonstrated that a configuration consisting of two $q^{Q}$-plates and $q^{H}$-plate, invariant under permutations of their ordering and constrained to identical topological charges of the constituent elements, spans the entire SU(2) group. This result, derived within a modified Euler-angle parameterization, establishes these configurations as the \textit{universal} SU(2) gadget for the HOPS.\\
\indent
A $q$-plate is a birefringent optical element, a member of the SU(2) family, characterized by a spatially varying fast-axis orientation across the transverse plane \cite{marrucci2006optical, machavariani2008spatially, slussarenko2011tunable, cardano2012polarization, marrucci2013q, ji2016meta, delaney2017arithmetic, qu2017plasma, kadiri2019wavelength, rubano2019q, ma2022generation, hakobyan2025q}. Depending on the imparted retardance between orthogonal polarization components, it operates as a $q^{Q}$-plate or $q^{H}$-plate, corresponding to retardances of $\pi/2$ and $\pi$, respectively. Each $q$-plate is defined by its topological charge $q$, denoting the number of complete rotations of the fast axis over one full azimuthal cycle, with $q$ admitting integer or half-integer values. Recently, a new type of $q$-plate has been introduced, where alongwith the azimuthally varying fast-axis orientation, the retardance also varies radially across the transverse plane \cite{hakobyan2025q}.\\
\indent
The objective of this work is to demonstrate systematic and controlled navigation on the HOPS via the realization of the SU(2) gadget within the theoretical framework of the effective waveplate formalism. In particular, we establish point-to-point navigation on the HOPS, where the offset angle of the constituent $q$-plates serves as the fundamental control parameter governing the evolution. This investigation provides a comprehensive account of polarization trajectories across the HOPS enabled by the effective SU(2) gadget. Since both the HOPS and the $q$-plate inherently possess topological attributes, we begin by presenting the essential background on beam topology, the associated topological spheres, and the role of structured polarizing elements, before advancing to the central results of this study.\\
\indent
The paper is structured as follows: In Section \ref{section_02}, we present the necessary formalism to ensure the completeness of the paper. This includes discussions on the HOPS, the $q$-plate and the concept of topological index space in polarization optics. Section \ref{section_03} focuses on the mathematical formulation of the effective SU(2) gadget, followed by a discussion on the effective waveplate. In Section \ref{section_04}, we present both the mathematical and geometrical descriptions of the transformation of the HOPS beam through the SU(2) gadget.  This section also highlights an important aspect, the radial symmetry in the topological index one space, along with its implications for the holonomic polarization evolution in the same space. The paper culminates in Section \ref{section9}.
\section{Formalism}
\label{section_02}
\subsection {Hgher-order Poincar\'{e} sphere}
\label{section2}
Mathematically, the HOPS beams are expressed as the superposition of two beams carrying equal but opposite OAM in orthogonal spin states as \cite{ruchi2020phase} 
\begin{equation}
    |\psi_{\ell}\rangle = \psi_{R}\ket{R_{\ell}}+\psi_{L}|\ket{L_{\ell}},
    \label{eqn_psi}
\end{equation}
where $|R_{\ell}\rangle$ and $|L_{\ell}\rangle$ are the orthogonal basis states embedded with optical vortex of topological charge $-\ell$ and $\ell$ respectively. The PH index of the
HOPS beam is half the difference between the topological charges of the vortex corresponding to $|L_{\ell}\rangle$ and $|R_{\ell}\rangle$, and hence $\eta=\ell$. The term $\psi_{R} = \bra{R_{\ell}}\psi_{\ell}\rangle$ and $\psi_{L}=\bra{L_{\ell}}\psi_{l}\rangle$ are the complex amplitudes given by 
\begin{align}
    \psi_{R} &= \cos{\left(\frac{\pi}{4}-\chi^{(\eta)}\right)}e^{-i\gamma^{(\eta)}}, \\
    \psi_{L} &= \sin{\left(\frac{\pi}{4}-\chi^{(\eta)}\right)}e^{i\gamma^{(\eta)}},
\end{align}
respectively, where $2\gamma^{(\eta)}$ and $2\chi^{(\eta)}$ represent the longitude and latitude coordinates on the HOPS, respectively. For $\ell=0$, Eq. (\ref{eqn_psi}) describes a plane wave with homogeneous polarization, reducing the HOPS to the standard PS, with longitude and latitude coordinates $2\chi^{(0)}$ and $2\gamma^{(0)}$, which correspond to the azimuth and ellipticity, respectively. For $\ell\neq 0$, $2\chi^{(\eta)}$ and $2\gamma^{(\eta)}$ correspond to ellipticity and Pancharatnam phase. On the HOPS, the equator is occupied with cylindrical vector beams (including radially and azimuthally polarized), where $|\psi_R| = |\psi_L|$, and non-equatorial points correspond to $|\psi_R| \neq |\psi_L|$.\\
\indent
The Stokes parameters (SPs) corresponding to the singular beam expressed in Eq. (\ref{eqn_psi}) are defined as $S_{0}^{(\eta)} = |\psi_R|^2 + |\psi_L|^2$, $S_{1}^{(\eta)} = 2 \, \text{Re}[\psi_R \psi_L^*]$, $S_{2}^{(\eta)} = 2 \, \text{Im}[\psi_R \psi_L^*]$ and $S_{3}^{(\eta)} = |\psi_R|^2 - |\psi_L|^2$.
These SPs are used to craft the HOPS for index $\eta$, as shown in Fig. \ref{sphere_01} for $\eta = -2$, $-1$, $0$, $+1$ and $+2$. These parameters satisfy the constraint equation $(S_{0}^{(\eta)})^2 = (S_{1}^{(\eta)})^2 + (S_{2}^{(\eta)})^2 + (S_{3}^{(\eta)})^2$. The coordinates $2\gamma^{(\eta)}$ and $2\chi^{(\eta)}$ in terms of Stokes parameters are expressed as $2\gamma^{(\eta)} = \tan^{-1}(S_{2}^{(\eta)} / S_{1}^{(\eta)})$ and $2\chi^{(\eta)} = \sin^{-1}(S_{3}^{(\eta)} / S_{0}^{(\eta)})$, respectively. For $\eta=l=0$, these higher-order SPs reduce to the standard SPs associated with the plane wave. 
\subsection{Spatially varying $q$-plate}
\label{section3}
The well-known QWP and HWP are SU(2) birefringent media characterized by a uniform retardance $\delta$ and a fast axis orientation $\alpha$ in the transverse plane, making them homogeneous waveplates. When one or both of these defining parameters (retardance or fast axis orientation) vary spatially, the waveplate is referred to as an inhomogeneous waveplate. A fundamental example of an inhomogeneous waveplate is the $q$-plate \cite{marrucci2006optical, machavariani2008spatially, slussarenko2011tunable, cardano2012polarization, marrucci2013q, ji2016meta, delaney2017arithmetic, qu2017plasma, kadiri2019wavelength, rubano2019q, ma2022generation, hakobyan2025q}, in which the fast axis orientation varies azimuthally in a systematic manner within the plane of the plate. This variation is mathematically given by $\alpha(\phi) = q\phi + \alpha_0$, where $q$ represents the topological charge and $\alpha_0$ is the offset angle. The topological charge $q$ specifies how much the fast axis rotates over one full azimuthal cycle and the offset angle $\alpha_{0}$ defines the fast axis orientation measured from a specified reference axis. The Jones matrix of the $q$-plate of retardance $\delta$ and fast axis $\alpha(\phi)$ is expressed as
\begin{widetext}
\begin{equation}
M(\delta, \alpha(\phi))= 
\begin{bmatrix}
        \cos \frac{\delta}{2} + i\sin\frac{\delta}{2} \cos 2 \alpha(\phi) & i\sin\frac{\delta}{2} \sin 2 \alpha(\phi) \\[10pt]
        i\sin\frac{\delta}{2} \sin 2 \alpha(\phi) & \cos \frac{\delta}{2} - i\sin\frac{\delta}{2} \cos 2 \alpha(\phi)
\end{bmatrix} \in \text{SU(2)}.
\label{matrix01}
\end{equation}
\end{widetext}
The Jones matrix of the $q$-plate features some properties as given below:
\begin{itemize}
    \item $M_{11}(\delta, \alpha(\phi))=[M_{22}(\delta, \alpha(\phi))]^{*}$
    \item $M_{12}(\delta, \alpha(\phi))=M_{21}(\delta, \alpha(\phi))$
    \item $M(\delta, \alpha(\phi)) = [M(\delta, \alpha(\phi))]^{T}$
    \item $\texttt{det}[M(\delta, \alpha(\phi))]=1$
    \item $[M(\delta, \alpha(\phi)]^{\dagger}M(\delta. \alpha(\phi))=\mathbf{I}$
\end{itemize}
The last two characteristics make this matrix an SU(2) matrix. Additionally, the off-diagonal elements are purely imaginary. Furthermore, if the Jones matrix of a $q$-plate is known, we can extract the information about the retardance $\delta$ and the fast axis orientation $\alpha(\phi)$ using the following expressions:
\begin{align}
\delta &= 2\cos^{-1}\left[\frac{1}{2}\texttt{Tr}[M(\delta, \alpha(\phi))]\right],
\label{delta}\\
\alpha (\phi) &= \frac{1}{2}\tan^{-1}\left[\frac{\texttt{Im}[M_{12}(\delta, \alpha(\phi))]}{\texttt{Im}[M_{11}(\delta, \alpha(\phi))]}\right].
\label{alpha}
\end{align}
\subsection{Topological index space in polarization optics}
\label{topological}
A topological index space in polarization optics was recently introduced \cite{umar2025holonomically}. In this space or world, each family comprises two member types: the topological sphere and the corresponding waveplates (also including their combinations). These waveplates and/or their combinations are selected in such a way that they facilitate holonomic SU(2) polarization evolution on the corresponding topological sphere. Consequently, in each family, the order of the sphere should be identical to the topological charge of the waveplate $\eta=q$. For example, the HWP, QWP, and their combinations like as QQ, QH, HQ, HH, SU(2) gadgets (QQH, QHQ, HQQ) and others, perform holonomic polarization evolution on the standard Poincaré sphere. In this case, the order of the sphere is $\eta=0$ and the topological charge of the waveplate is $q=0$, meaning these elements belong to the index zero space. \\
\indent
Next, the $q^{Q}$-plate, $q^{H}$-plate, higher-order SU(2) gadgets such as $q^{Q}q^{Q}q^{H}$, $q^{Q}q^{H}q^{Q}$ and $q^{H}q^{Q}q^{Q}$ each with a topological charge of $q=1$, perform holonomic transformations on the HOPS of order $\eta=1$. Therefore, both the sphere and the aforementioned $q$-plates and combinations are members of the index one space. In general, the index $\eta$ (or $q$) space consists of the topological sphere of order $\eta$ and the corresponding waveplates and their combinations with charge equal to the index. This equality $q=\eta$, called the holonomy condition, is necessary for holonomic polarization evolution to occur \cite{umar2025holonomically, umar2025mathematics}.
\section{Effective SU(2) gadget}
\label{section_03}
\subsection{Effective waveplate}
\label{section4}
The concept of an effective waveplate \cite{kadiri2019wavelength, umar2025mathematics} arises when multiple waveplates, whether homogeneous or inhomogeneous, are aligned coaxially, such that under specific conditions, the resulting system mimics the optical behavior of a single, effective waveplate having effective retardance $\delta_{e}$ and effective fast axis orientation $\alpha_{e}$. We have recently provided a detailed discussion of the combination of three $q$-plates within the framework of effective waveplates, as reported in \cite{umar2025mathematics}. To ensure the completeness of this paper, a discussion on the effective waveplate is necessary. 
Revisit the SU(2) Jones matrix expressed in Eq. (\ref{matrix01}) which admits the form
\begin{equation}
M(\delta, \alpha(\phi))= 
\begin{bmatrix}
        \mathbbmss{A}+i\mathbbmss{B} & i\mathbbmss{D} \\[8pt]
        i\mathbbmss{D} & \mathbbmss{A}-i\mathbbmss{B}
\end{bmatrix},
\label{matrix02}
\end{equation}
where, $\mathbbmss{A}$, $\mathbbmss{B}$ and $\mathbbmss{D}$ are the function $f(\delta, \alpha(\phi))$. The diagonal elements are the complex conjugates, while the off-diagonal elements are identical and purely imaginary. These constraints define the properties that a Jones matrix must satisfy.\\
\indent
Consider a system of $n$ $q$-plates with retardances $\delta_1$, $\delta_2$, $\delta_3$, $\dots$, $\delta_n$ and fast axis orientations $\alpha_1(\phi)$, $\alpha_2(\phi)$, $\alpha_{3}(\phi)$, $\dots$, $\alpha_n(\phi)$ arranged coaxially. Using the composition property of matrices, the entire setup can be expressed mathematically as:
\begin{multline}
\mathbb{M} = M(\delta_n, \alpha_n(\phi)) \cdot M(\delta_{n-1}, \alpha_{n-1}(\phi)) \cdot \dots \\
\cdots \cdot M(\delta_3, \alpha_3(\phi)) \cdot M(\delta_2, \alpha_2(\phi)) \cdot M(\delta_1, \alpha_1(\phi)).
\label{composition_matrix}
\end{multline}
After multiplying the Jones matrices, the resulting matrix $\mathbb{M}$ may not necessarily match the form in Eq. (\ref{matrix02}). Instead, it can yield a matrix with real components in the off-diagonal elements. The matrix in Eq. (\ref{composition_matrix}) is expressed as
\begin{equation} 
\mathbb{M}= 
\begin{bmatrix} \mathbbmss{A}+i\mathbbmss{B} & \mathbbmss{C}+i\mathbbmss{D} \\[14pt] -\mathbbmss{C}+i\mathbbmss{D} & \mathbbmss{A}-i\mathbbmss{B} 
\end{bmatrix}. 
\label{matrix04} 
\end{equation}
The resultant matrix for a combination of $q$-plates differs from that of a single $q$-plate due to the presence of the real component $\mathbbmss{C}$ in the off-diagonal elements. To replicate the behavior of a single effective $q$-plate, the condition $\mathbbmss{C} = 0$ must be satisfied. This condition serves as the necessary constraint for the waveplates to function effectively as a single $q$-plate. The matrix elements $\mathbbmss{A}$, $\mathbbmss{B}$, $\mathbbmss{C}$ and $\mathbbmss{D}$ having the functional form as $f(\delta_1, \delta_2, \delta_3, \dots, \delta_n, \alpha_1, \alpha_2, \alpha_3, \dots, \alpha_n)$. Thus, by appropriately selecting the waveplate parameters, the system can be configured such that $\mathbbmss{C}$ vanishes, leading to the effective waveplate form $\mathbb{M} = M(\delta_e, \alpha_e(\phi))$.
\subsection{Three $\textbf{\textit{q}}$-plates}
\label{section5}
Consider three $q$-plates with retardances $\delta_1$, $\delta_2$ and $\delta_3$, and fast axis orientations $\alpha_j(\phi) = q_j \phi + \alpha_{0j}$, where $j = 1$, $2$, $3$, arranged coaxially. The combined effect of this configuration can be described using the matrix composition property, as follows
\begin{equation}
\begin{aligned}
\mathbb{M} & = M(\delta_{3}, \alpha_{3}(\phi)) \cdot M(\delta_{2}, \alpha_{2}(\phi)) \cdot M(\delta_{1}, \alpha_{1}(\phi)) \\[8pt]
        & =
\begin{bmatrix}
\mathbbmss{A}_{q}+i\mathbbmss{B}_{q} & \mathbbmss{C}_{q}+i\mathbbmss{D}_{q} \\[8pt]
-\mathbbmss{C}_{q}+i\mathbbmss{D}_{q} & \mathbbmss{A}_{q}-i\mathbbmss{B}_{q}
\end{bmatrix} \in \text{SU}(2).
\end{aligned}
\label{matrix05_06}
\end{equation}
This matrix respects the form presented in Eq. (\ref{matrix04}), where the term $\mathbbmss{C}_{q}$ is given by
\begin{multline}
    \mathbbmss{C}_{q} = -\cos\frac{\delta_{1}}{2}\sin\frac{\delta_{2}}{2}\sin\frac{\delta_{3}}{2}\sin2[\alpha_{2}(\phi)-\alpha_{3}(\phi)]\\-\cos\frac{\delta_{2}}{2}\sin\frac{\delta_{1}}{2}\sin\frac{\delta_{3}}{2}\sin2[\alpha_{1}(\phi)-\alpha_{3}(\phi)]\\-\cos\frac{\delta_{3}}{2}\sin\frac{\delta_{1}}{2}\sin\frac{\delta_{2}}{2}\sin2[\alpha_{1}(\phi)-\alpha_{2}(\phi)],
    \label{m12}
\end{multline}
The presence of a nonzero real component in the off-diagonal elements, $\mathbbmss{C}_q \neq 0$, implies that the configuration cannot be regarded as a single effective $q$-plate unless $\mathbbmss{C}_q$ vanishes. Furthermore, for the system of three $q$-plates to function as a single effective waveplate, two conditions must be satisfied: the fast axis of the effective $q$-plate must lie in the plane of the plate and exhibit azimuthal dependence similar to that of a single $q$-plate and the effective $q$-plate must maintain uniform spatial retardance across its transverse plane.\\
\indent
The general analytical solution for $\mathbbmss{C}_q = 0$ is mathematically complex. The central theme of this paper is to discuss the SU(2) gadget within the framework of an effective waveplate. The SU(2) gadgets are the $q^{Q}q^{Q}q^{H}$, $q^{Q}q^{H}q^{Q}$ and $q^{H}q^{Q}q^{Q}$ configurations \cite{umar2025su2gadget}. We will proceed only by focusing on these specific configurations.
\subsection{$q^{Q}q^{H}q^{Q}$ configuration}
In this configuration, the $q^{H}$-plate is sandwiched between two $q^{Q}$-plates. For such an arrangement, the retardances are $\delta_{1} = \delta_{3} = \pi/2$ and $\delta_{2} = \pi$. Under these conditions, the corresponding term $\mathbbmss{C}_{q}$ is expressed as
\begin{equation}
    \mathbbmss{C}_{q}= \sin\alpha_{31}(\phi)\cos\left[\alpha^{+}_{13}(\phi)-2\alpha_{2}(\phi)\right],
\end{equation}
where, $\alpha_{xy}(\phi) = \alpha_{x}(\phi) - \alpha_{y}(\phi)$ and $\alpha^{+}_{xy}(\phi) = \alpha_{x}(\phi) + \alpha_{y}(\phi)$. To make $\mathbbmss{C}_q = 0$, one of two conditions must hold: (i) $|\alpha^{+}_{13}(\phi) - 2\alpha_{2}(\phi)| = \pi/2$ or (ii) $\alpha_{31}(\phi) = 0 \Rightarrow \alpha_{1}(\phi) = \alpha_{3}(\phi)$. Applying the first condition to the matrix in Eq. (\ref{matrix05_06}) and evaluating the effective parameters using Eqs. (\ref{delta}) and (\ref{alpha}), we obtain $\delta_{e} = \pi$ and $\alpha_{e}(\phi) = \alpha_{2}(\phi)$. In this case, the system behaves as a $q^{H}$-plate with fast axis orientation as $\alpha_{2}(\phi)$.\\
\indent
Next, applying the first condition, $\alpha_{1}(\phi) = \alpha_{3}(\phi)$, in the matrix expressed in Eq. (\ref{matrix05_06}) and calculating the effective parameters, we arrive at
\begin{equation}
\delta_{e}
= \pi + 2\sin^{-1}\!\Big[\cos2[(q_{1}-q_{2})\phi + (\alpha_{01}-\alpha_{02})]\Big],
\label{rhoe01}
\end{equation}
\begin{equation}
\alpha_{e}(\phi)
= -\frac{1}{2}\tan^{-1}\big[\cot(2\alpha_{1}(\phi))\big]
= \alpha_{1}(\phi) - \frac{\pi}{4}.
\label{alphae01}
\end{equation}
Recall that, by definition, an effective $q$-plate requires a uniform retardance. However, the effective retardance derived above depends on the azimuthal angle $\phi$, which violates this condition. To ensure uniform retardance, it is necessary that $q_1 = q_2$, meaning the topological charges of the $q$-plates must be equal. Under this condition, the effective retardance becomes
\begin{align}
\delta_{e} &= \pi + 2\sin^{-1}\Big[\cos 2(\alpha_{01} - \alpha_{02})\Big] \nonumber \\
             &= 2\pi - 4(\alpha_{Q} - \alpha_{H})=2\pi-4\Delta\alpha. 
             \label{varretardance}
\end{align}
In the above expression, we define $\alpha_{01} = \alpha_{Q}$ as the offset angle of the $q^{Q}$-plate, $\alpha_{02} = \alpha_{H}$ as that of the $q^{H}$-plate, and $\Delta\alpha = \alpha_{Q} - \alpha_{H}$. A particularly noteworthy outcome of this formulation is the tunability of the effective retardance, governed entirely by the relative offset angle. Notably, $\delta_{e} = 2\pi$ for $\Delta\alpha = 0$, and $\delta_{e} = 0$ for $\Delta\alpha = \pi/2$. Thus, by varying $\Delta\alpha$ from $\pi/2$ to $0$, the effective retardance can be continuously tuned from $0$ to $2\pi$. Eq. (\ref{alphae01}) can further be expressed as 
\begin{equation}
     \alpha_{e}(\phi) = q\phi + \underbrace{\left(\alpha_{Q} - \pi/4\right)}_{\substack{\text{Effective offset} \\ \text{angle}}}.
     \label{varoffset}
\end{equation}
It should be noted that, in this formulation, all $q$-plates share the same topological charge, i.e., $q_{1} = q_{2} = q_{3} = q$, with $\alpha_{01} = \alpha_{03} = \alpha_{Q}$ and $\alpha_{02} = \alpha_{H}$. Consequently, the $q^{Q}q^{H}q^{Q}$ configuration effectively behaves as a single $q$-plate of charge $q$, with effective retardance $\delta_{e}$ (Eq. (\ref{varretardance})) and an effective offset angle of $\alpha_{Q} - \pi/4$.
\subsection{$q^{Q}q^{Q}q^{H}$ configuration}
In this arrangement $\delta_{1}=\delta_{2}=\pi/2$ and $\delta_{3}=\pi$. For this setup, the $\mathbbmss{C}_{q}$ term is expressed as
\begin{equation}
    \mathbbmss{C}_{q}=\cos{\alpha_{12}(\phi)}\sin{\left[2\alpha_{3}(\phi)-\alpha_{12}^{+}(\phi)\right]}.
\end{equation}
For this configuration to function as an effective waveplate, the condition $\mathbbmss{C}{q} = 0$ must be satisfied. This requirement can be fulfilled under either of the following conditions: (i) $|\alpha_{1}(\phi) - \alpha_{2}(\phi)| = \pi/2$, or (ii) $2\alpha_{3}(\phi) = \alpha_{12}^{+}(\phi)$. Under the first condition, the effective parameters are obtained as $\delta_{e} = \pi$ and $\alpha_{e}(\phi) = \alpha_{3}(\phi)$, showing that the configuration behaves effectively as a $q^{H}$-plate. The second condition yields the effective retardance and fast axis orientation identical to those given in Eqs.~(\ref{varretardance}) and (\ref{varoffset}), respectively. Here too, imposing $q_{1} = q_{2} = q_{3} = q$ guarantees that the effective retardance $\delta_{e}$ is independent of the azimuthal angle $\phi$. 
\subsection{$q^{H}q^{Q}q^{Q}$ configuration}
In this configuration, $\delta_{1} = \pi$ while $\delta_{2}$ and $\delta_{3} = \pi/2$. For this setup, the corresponding $\mathbbmss{C}_{q}$ term is given by
\begin{equation}
    \mathbbmss{C}_{q}=\cos{\alpha_{23}(\phi)}\sin{\left[\alpha_{23}^{+}(\phi)-2\alpha_{1}(\phi)\right]}.
\end{equation}
Again, for this configuration to function as an effective waveplate, the condition $\mathbbmss{C}_{q} = 0$ must be satisfied, which can be achieved if either (i) $|\alpha_{2}(\phi) - \alpha_{3}(\phi)| = \pi/2$ or (ii) $\alpha_{23}^{+}(\phi) = 2\alpha_{1}(\phi)$. Under the first condition, the effective parameters are $\delta_{e} = \pi$ and $\alpha_{e}(\phi) = \alpha_{3}(\phi)$, showing that the configuration effectively behaves as a $q^{H}$-plate. Next, under the second condition, the effective parameters are the same as those given in Eqs. (\ref{varretardance}) and (\ref{varoffset}).\\
\indent
A central and significant outcome is that all three configurations, $q^{Q}q^{Q}q^{H}$, $q^{Q}q^{H}q^{Q}$ and $q^{H}q^{Q}q^{Q}$, when subjected to the appropriate constraints, reduce to an effective single $q$-plate of topological charge $q$ with an offset angle of $\alpha_{Q} - \pi/4$. Crucially, this effective $q$-plate exhibits tunable retardance, as given by Eq. (\ref{varretardance}), governed directly by variations in $\Delta\alpha$. In general, these configurations constitute the SU(2) gadget for the HOPS and, when interpreted through the framework of effective waveplates, can therefore be referred to as the effective SU(2) gadget in this context.
\section{Polarization evolution: Role of the effective SU(2) gadget}
\label{section_04}
In the preceding section, it was established that a configuration comprising two $q^{Q}$-plates and one $q^{H}$-plate, invariant under permutations of their ordering, exhibits the nontrivial property of tunable retardance when subjected to specific constraints. In the following section, this framework is extended to analyze the transformation of beams on the HOPS through the action of these gadgets.
\subsection{Mathematical description}
When the HOPS beam defined in Eq. (\ref{eqn_psi}) is passed through these gadgets, the output beam, under the holonomy condition $\ell = \eta = q$ \cite{umar2025holonomically}, is given by 
\begin{equation}
    |\psi^{'}_{\ell}\rangle = \psi^{'}_{R}\ket{R_{\ell}}+\psi^{'}_{L}|\ket{L_{\ell}},
    \label{eqn_psi01}
\end{equation}
where, $\psi^{'}_{R}$ and $\psi^{'}_{L}$ are the complex amplitudes and is expressed as
\begin{align}
    \psi^{'}_{R} &= \frac{1}{\sqrt{2}}\left[i(\cos{\chi^{(\eta)}}-\sin{\chi^{(\eta)}})\sin{\frac{\delta_{e}}{2}}e^{-i\left(2\alpha_{Q}-\gamma^{(\eta)}-\frac{\pi}{2}\right)} \right. \nonumber \\
    &\quad \left. + (\cos{\chi^{(\eta)}}+\sin{\chi^{(\eta)}})\cos{\frac{\delta_{e}}{2}}e^{-i\gamma^{(\eta)}} \right],
\end{align}
\begin{align}
    \psi^{'}_{L} &= \frac{1}{\sqrt{2}}\left[i(\cos{\chi^{(\eta)}}+\sin{\chi^{(\eta)}})\sin{\frac{\delta_{e}}{2}}e^{i\left(2\alpha_{Q}-\gamma^{(\eta)}-\frac{\pi}{2}\right)} \right. \nonumber \\
    &\quad \left. + (\cos{\chi^{(\eta)}}-\sin{\chi^{(\eta)}})\cos{\frac{\delta_{e}}{2}}e^{i\gamma^{(\eta)}} \right],
\end{align}
where, $\delta_{e}$ represents the effective retardance of the gadget. It is evident that the output beam retains the same functional form as Eq.~(\ref{eqn_psi}), as it constitutes a superposition of right- and left-circularly polarized vortex beams with opposite topological charges and the topological charge is same as the input beam, a consequence of enforcing the holonomy condition. Consequently, the input and output beams lie on the same sphere.
\subsection{Geometrical (rotational) description}
\begin{figure*}[t]
\begin{center}
\setlength{\fboxsep}{10pt}
\tikzset{dottedbox/.style={draw, color=red, line width=0.5mm, dotted, rounded corners=10pt, inner sep=15pt, outer sep=0pt
}}
\begin{tikzpicture}
\node[dottedbox] {\includegraphics[scale=0.27]{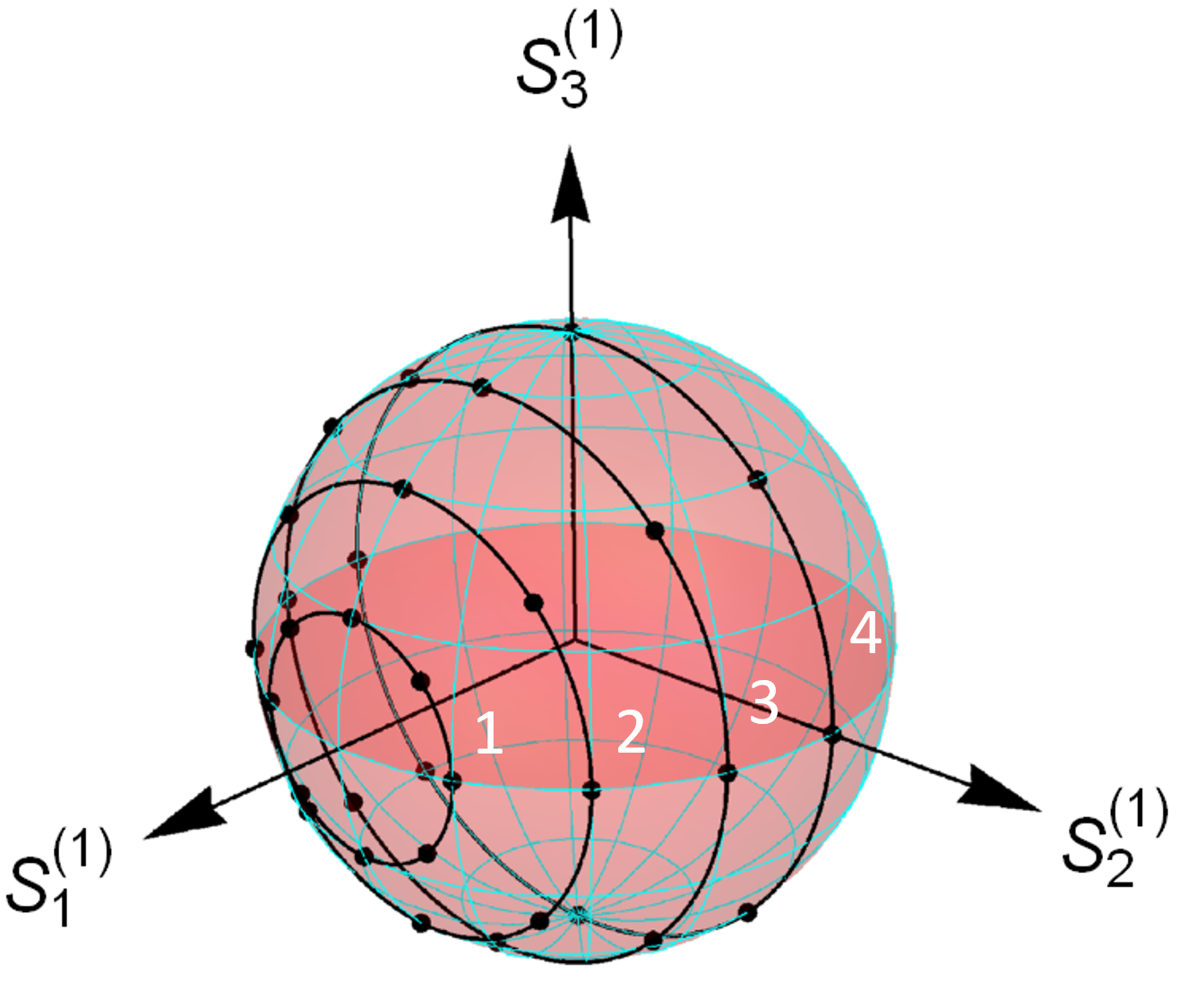} \textcolor{black}{(a)} \hspace{40pt}
\includegraphics[scale=0.27]{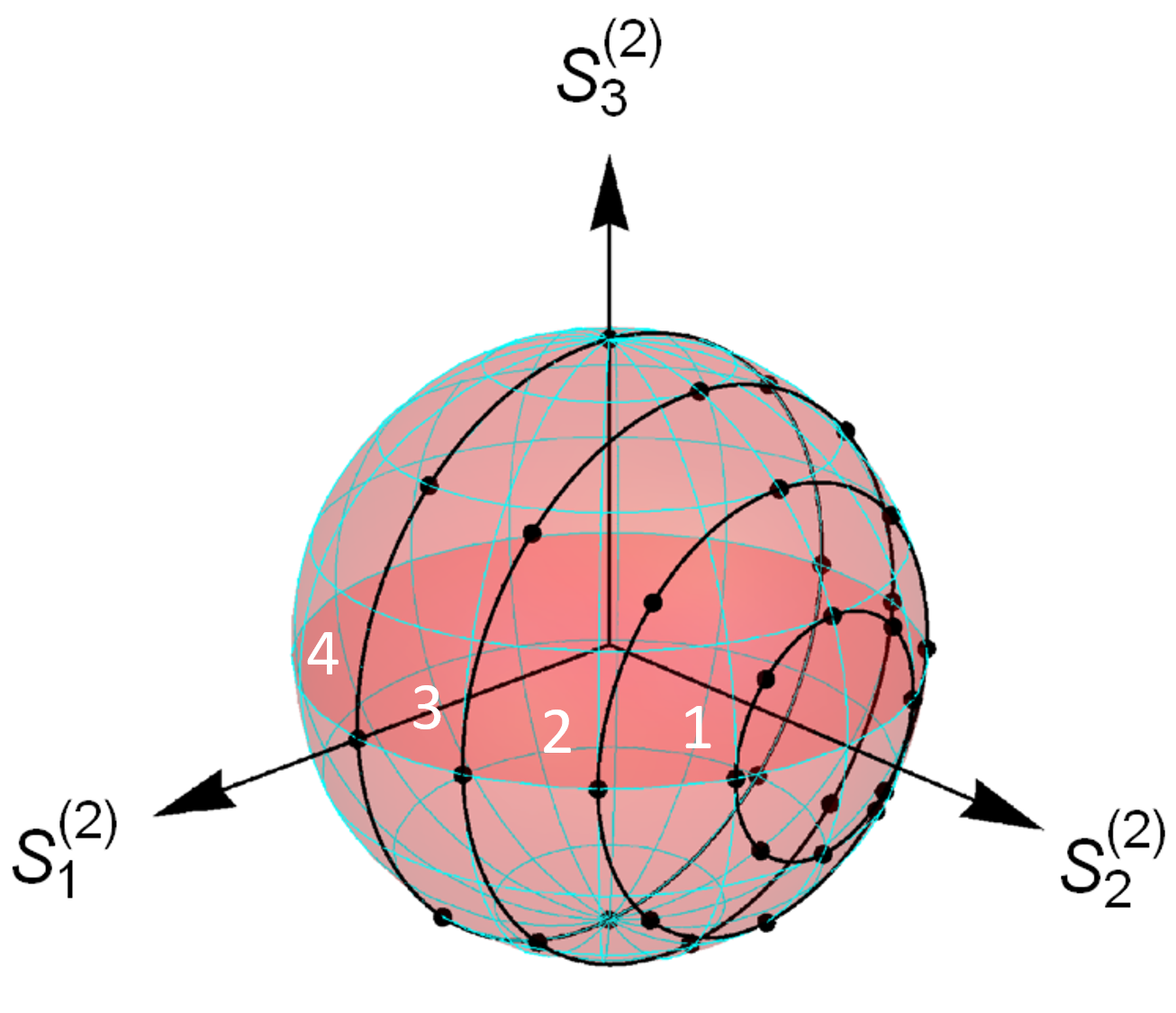} \textcolor{black}{(b)}
};
\end{tikzpicture}
\caption{(Color online). Figure showing (a) the HOPS of order $\eta=1$ with four circles lying on its surface around the $S_{1}^{(1)}$-axis, labeled $1$–$4$, and (b) the HOPS of order $\eta=2$ with four circles lying on its surface around the $S_{2}^{(2)}$-axis, labeled $1$–$4$. Each circle contains eight points separated by an angular interval of $\pi/4$.}
\label{hops_01}
\end{center}
\end{figure*}

\begin{figure*}[p] 
\centering 
\vspace*{\fill} 
\includegraphics[width=0.85\textwidth, keepaspectratio]{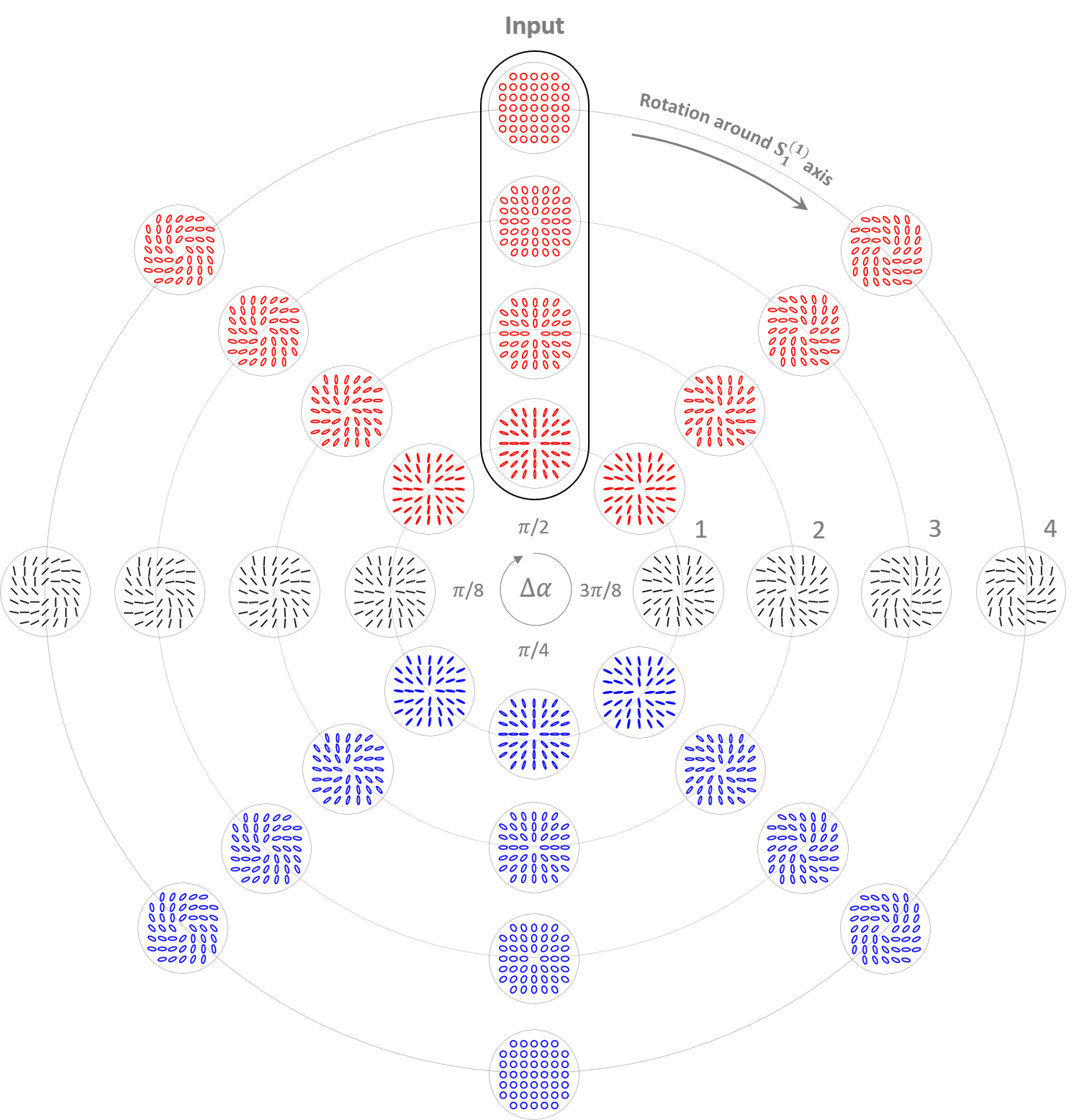} 
\caption{(Color online). Figure illustrating the polarization transformation on the HOPS of order $\eta = 1$ through the effective $q^{Q}q^{H}q^{Q}$ gadget. The input beam, chosen on the $S_{1}^{(1)}$–$S_{3}^{(1)}$ plane (shown here), undergoes a complete $2\pi$ rotation aroud the $S_{1}^{(1)}$ axis on the HOPS as $\Delta \alpha = \alpha_{Q} - \alpha_{H}$ is varied from $\pi/2$ to $0$. To achieve rotation around the $S_{1}^{(1)}$ axis, $\alpha_{Q}$ is set to $\pi/4$, effectively making the offset angle zero and aligning the rotation axis with the $S_{1}^{(1)}$ axis. For each point on the circles of the HOPS shown in Fig. \ref{hops_01}a, the corresponding HOPS beam is presented here.} 
\label{sphere_011} 
\vspace*{\fill} 
\end{figure*}

\begin{figure*}[p] 
\centering 
\vspace*{\fill} 
\includegraphics[width=0.85\textwidth, keepaspectratio]{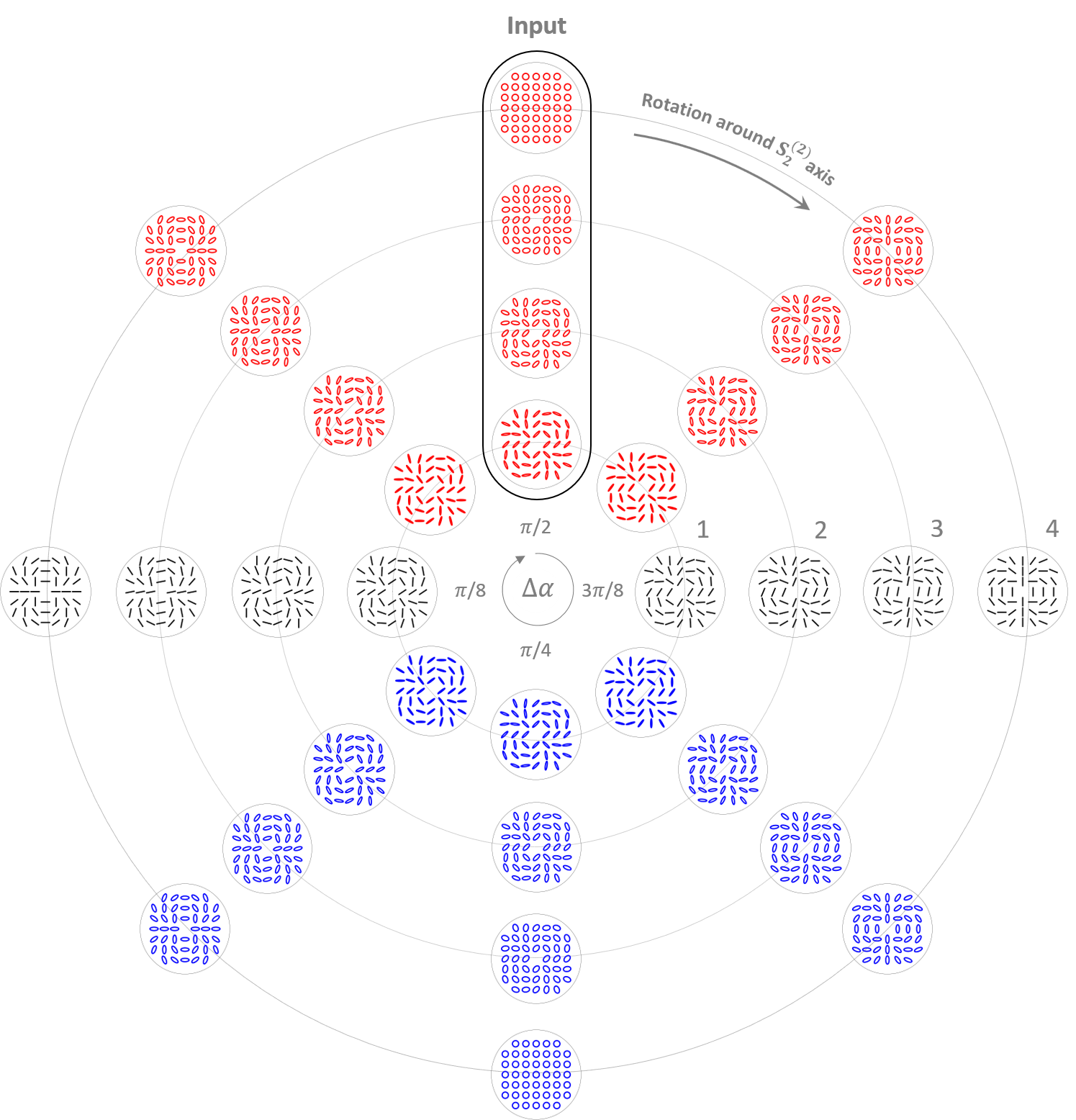} 
\caption{(Color online). Figure illustrating the polarization transformation on the HOPS of order $\eta = 2$ through the effective $q^{Q}q^{H}q^{Q}$ gadget. The input beam, chosen on the $S_{2}^{(2)}$–$S_{3}^{(2)}$ plane (shown here), undergoes a complete $2\pi$ rotation around the $S_{2}^{(2)}$ axis on the HOPS as $\Delta \alpha = \alpha_{Q} - \alpha_{H}$ is varied from $\pi/2$ to $0$. To achieve rotation around the $S_{2}^{(2)}$ axis, $\alpha_{Q}$ is set to $\pi/2$, making the effective offset angle $\pi/4$ and consequently aligning the rotation axis with the $S_{2}^{(2)}$ axis. For each point on the circles of the HOPS shown in Fig. \ref{hops_01}b, the corresponding HOPS beam is presented here.} 
\label{sphere_022} 
\vspace*{\fill} 
\end{figure*}

Before venturing into the study of polarization evolution on the HOPS using the SU(2) gadget within the effective waveplate framework, it is first necessary to discuss how, under the holonomy condition, such polarization transformations can be implemented using a $q$-plate. 
\begin{tcolorbox}[colback=myshade, colframe=black!60,
                  arc=10pt, boxrule=0.4pt]
The SU(2) transformation imparted by a homogeneous waveplate on a homogeneous beam maps to an SO(3) rotation on the standard Poincaré sphere. In this rotation, the rotation axis is decided by the fast axis orientation, while the retardance defines the rotation angle \cite{kumar2011polarization}. Analogously, the SU(2) transformation imparted by a $q$-plate on a HOPS beam of order $\eta = q$ manifests as an SO(3) rotation on the sphere. In this case, the rotation axis is determined by the offset angle of the $q$-plate, while the retardance specifies the rotation angle \cite{yao2023quantitative, umar20252}. The rotation axis lies in the equatorial plane of the sphere and is oriented at an angle twice the offset angle with respect to the $S_{1}^{(\eta)}$ axis in the Stokes space.
\end{tcolorbox}
The geometry illustrated in Fig. \ref{hops_01}a represents an $\textbf{S}^{2}$ HOPS of order $\eta = 1$. On the surface of the sphere, four circular trajectories are drawn around the $S_{1}^{(1)}$-axis, labeled 1 through 4. Each circle is discretized into eight equidistant points, separated by angular intervals of $\pi/4$. These circular trajectories are employed to visualize the transformation of HOPS beams as they evolve along these 
paths when propagated through the $q^{Q}q^{H}q^{Q}$ gadget under variation of the relative offset angle $\Delta \alpha$. For $\eta = 1$, the holonomic polarization transformations require a gadget with topological charge $q = 1$. Furthermore, the rotations on the HOPS are performed about the $S_{1}^{(1)}$-axis. To align the rotation axis with $S_{1}^{(1)}$-axis, the effective offset angle $\alpha_{Q} - \pi/4$ must vanish, which is satisfied when $\alpha_{Q} = \pi/4$. This requirement arises because, in the equatorial plane of the HOPS, the angle between the rotation axis and the $S_{1}^{(\eta)}$-axis is twice the effective offset angle.\\
\indent
Next, Fig. \ref{sphere_011} presents the projection of the four circles shown in Fig. \ref{hops_01}a, together with the corresponding polarization distributions at the designated points. The chosen input beam coordinates on the circles 1, 2, 3 and 4 are $(2\gamma^{(1)}, 2\chi^{(1)}) =$ $(0, \pi/8)$, $(0, \pi/4)$, $(0, 3\pi/8)$ and $(0, \pi/2)$, respectively. Consider circle 1, as shown in Figs. \ref{hops_01}a and \ref{sphere_011}, with the corresponding input beam. When this input beam is propagated through the $q^{Q}q^{H}q^{Q}$ gadget (with $q = 1$ and $\alpha_{Q} = \pi/4$) and the relative offset angle $\Delta\alpha_{QH} = \alpha_{Q} - \alpha_{H}$ is varied from $\pi/2$ to $0$, the resulting output HOPS beam traces the trajectory of circle 1. The output states corresponding to the variation of $\Delta \alpha$ at the step of $\pi/4$ are illustrated in Fig. \ref{sphere_011}. As $\Delta \alpha$ decreases from $\pi/2$ to $0$, the beam undergoes a complete $2\pi$ rotation around the $S_{1}^{(1)}$-axis. A similar explanation applies to circles $2$, $3$ and $4$, as shown in Figs. \ref{hops_01}a and \ref{sphere_011}. Table \ref{table_01} presents the tabulated values of $\alpha_H$ for a fixed $ \alpha_Q = \pi/4$, corresponding to the effective retardance $\delta_e$ over the range from $ 0 $ to $ 2\pi $ in steps of $ \pi/4 $. The corresponding values of $\Delta \alpha $ are also provided.\\
\begin{table}[h]
\centering
\setlength{\tabcolsep}{27.4pt}
\renewcommand{\arraystretch}{1.4} 
\begin{tabular}{|c|c|c|}
\hline
\textbf{$\bm{\textcolor{red}{\alpha_{H}}}$} & \textbf{$\bm{\textcolor{red}{\Delta\alpha}}$}& \textbf{$\bm{\textcolor{red}{\delta_{e}}}$} \\ \hline
$-\pi/8$ & $3\pi/8$ & $\pi/2$ \\ \hline
$-\pi/16$ & $5\pi/16$ & $3\pi/4$ \\ \hline
$0$  & $\pi/4$  & $\pi$ \\ \hline
$\pi/16$  & $3\pi/16$  & $5\pi/4$ \\ \hline
$\pi/8$ & $\pi/8$ & $3\pi/2$ \\ \hline
$3\pi/16$ & $\pi/16$ & $7\pi/4$ \\ \hline
$\pi/4$ & $0$ & $2\pi$ \\ \hline
\end{tabular}
\caption{Tabulated values of $ \alpha_H $ for a fixed $\alpha_Q = \pi/4$, corresponding to the effective retardance $ \delta_e $ over the range from $0$ to $2\pi$ in steps of $\pi/4$. The corresponding value of $ \Delta \alpha =\alpha_{Q}-\alpha_{H}$ is also provided.
}
\label{table_01}
\end{table}
\indent
Consider the geometry depicted in Fig. \ref{hops_01}b, which represents the HOPS of order $\eta=2$. Again, there are four circles on the HOPS, labeled as $1$, $2$, $3$ and $4$ around the $S_{2}^{(2)}$-axis, containing equidistant points separated by intervals of $\pi/4$. In this case, the input HOPS beam coordinates on the circles 1, 2, 3 and 4 are $(2\gamma^{(1)}, 2\chi^{(1)}) = (\pi/2, \pi/8)$, $(\pi/2, \pi/4)$, $(\pi/2, 3\pi/8)$ and $(\pi/2, \pi/2)$, respectively. Here, the circles are centered on the $S_{2}^{(2)}$-axis, and therefore this is the rotation axis, implying an angular separation of $\pi/2$ between the rotation axis and the $S_{1}^{(2)}$-axis. This implies that twice the effective offset angle satisfies $2(\alpha_{Q} - \pi/4) = \pi/2$, showing that the offset angle should be set $\alpha_{Q}$ at $\pi/2$. When these input beams are passed through the $q^{Q}q^{H}q^{Q}$ gadget ($q=1$ and $\alpha_{Q} = \pi/2$) and the relative offset angle $\Delta \alpha$ is varied from $\pi/2$ to $0$, the resulting output HOPS beams trace the trajectories of their respective circles, as illustrated in Figs. \ref{hops_01}b and \ref{sphere_022}. Fig. \ref{sphere_022} illustrates the output beams obtained by varying $\Delta \alpha$ in steps of $\pi/16$. The Table \ref{table_02} presents the values of $\alpha_H$ for $\alpha_Q = \pi/2$, which correspond to the effective retardance $ \delta_e $ across the range from $ 0 $ to $ 2\pi $ in increments of $\pi/4 $. The table also includes the corresponding values of $ \Delta \alpha$.
\clearpage
\begin{table}[h]
\centering
\setlength{\tabcolsep}{29pt}
\renewcommand{\arraystretch}{1.4} 
\begin{tabular}{|c|c|c|}
\hline
\textbf{$\bm{\textcolor{red}{\alpha_{H}}}$} & \textbf{$\bm{\textcolor{red}{\Delta\alpha}}$}& \textbf{$\bm{\textcolor{red}{\delta_{e}}}$} \\ \hline
$\pi/16$ & $7\pi/16$ & $\pi/4$ \\ \hline
$\pi/8$ & $3\pi/8$ & $\pi/2$ \\ \hline
$3\pi/16$ & $5\pi/16$ & $3\pi/4$ \\ \hline
$\pi/4$  & $\pi/4$  & $\pi$ \\ \hline
$5\pi/16$  & $3\pi/16$  & $5\pi/4$ \\ \hline
$3\pi/8$ & $\pi/8$ & $3\pi/2$ \\ \hline
$7\pi/16$ & $\pi/16$ & $7\pi/4$ \\ \hline
$\pi/2$ & $0$ & $2\pi$ \\ \hline
\end{tabular}
\caption{Tabulated values of $ \alpha_H $ for a fixed $ \alpha_Q = \pi/2 $, corresponding to the effective retardance $ \delta_e $ over the range from $ 0 $ to $ 2\pi $ in increments of $ \pi/4 $. The associated values of $ \Delta \alpha =\alpha_{Q}-\alpha_{H}$ are also provided.
}
\label{table_02}
\end{table}
\indent
These results show that, by choosing $\alpha_{Q}$ as per the requirement and systematically varying the relative offset angle $\Delta \alpha$, any arbitrary polarization state can be realized. To substantiate our result, circular trajectories on the sphere are first delineated, and it is then demonstrated that movement along these trajectories is attainable through the effective SU(2) gadget. More generally, polarization evolution on the HOPS can be described as a rotation by an angle about a rotation axis that always lies in the equatorial plane. The concept of an effective SU(2) gadget fits here, providing a means for controlled navigation on the HOPS. In the discussion above, we have illustrated the geometric picture of polarization evolution using the $q^{Q}q^{H}q^{Q}$ gadget, however, the same explanation applies equally to the other two configurations, namely effective $q^{Q}q^{Q}q^{H}$ and $q^{H}q^{Q}q^{Q}$ gadgets.  
\subsection{Implication of radial symmetry in topological index one space}
\label{section_06}
\label{section_05}
In Section \ref{topological}, the concept of the topological index space, recently introduced in \cite{umar2025holonomically}, is presented. The members of the index one family in this space are the HOPS of order $\eta = 1$, together with the structured optical elements of topological charge $q = 1$. It is important to note that both beams and optical elements of order/charge one exhibit radial symmetry in their textures, implying that the entire index one space possesses radial symmetry. As shown in the preceding discussion, controlled evolution on the sphere can be achieved through variation of $\alpha_{Q}$ and $\Delta \alpha$. However, for gadgets composed of $q=1$ plates, such variation cannot be realized by merely rotating the plates, since a $q=1$ plate retains its radial symmetry with respect to any reference axis and thus its texture remains unchanged under rotation. In contrast, for gadgets constructed with $q \neq 1$, variation in $\alpha_{Q}$ and $\Delta \alpha$ can indeed be achieved through physical rotation of the constituent plates. The practical implementation of such variations in the gadgets can be achieved through advanced photonic platforms, including metasurfaces, metamaterials and liquid crystal–based technologies.
\section{Conclusion}
\label{section9}
In this paper, we have presented a controlled holonomic polarization walk on the HOPS using SU(2) gadgets formulated under the concept of an effective waveplate. Although the SU(2) gadget has previously been reported in \cite{umar2025su2gadget} through a modified Euler-angle parameterization, the present work differs in that the gadget is examined within the framework of the effective waveplate, and in this context, it may be referred to as an effective SU(2) gadget. Any arbitrary polarization evolution on the HOPS can be described as a rotation by a certain angle about a rotation axis that always lies in the equatorial plane. The effective waveplate formulation provides a systematic means of determining both the orientation of the rotation axis within the equatorial plane and the magnitude of the rotation angle. Specifically, the effective offset angle of the gadget defines the orientation of the axis, while the effective retardance sets the rotation angle. The effective waveplate paradigm exhibits the property of tunable retardance, continuously adjustable from $0$ to $2\pi$ as the relative offset angle is varied from $\pi/2$ to $0$ for a given $\alpha_{Q}$. Leveraging this fact, we have demonstrated a holonomic walk on the HOPS. It is also important to note that the concept of the (effective) SU(2) gadget is universal, applying to both positive and negative order spheres, provided that the holonomy condition is satisfied. The theoretical foundation of the effective SU(2) gadget has been established, and this study points toward potential applications in realizing its experimental textures. The findings bear direct relevance to the deterministic control and engineering of structured light, encompassing polarization singularities, vector vortex beams, and topological optical fields. Moreover, these findings are deeply linked to current research frontiers in singular optics, spin–orbit photonics, and quantum information science, where the controlled realization of polarization evolution within higher-order state spaces represents a challenge of both fundamental interest and technological relevance.
\begin{acknowledgments}
MU gratefully acknowledges the institute fellowship support from IIT Delhi and extends heartfelt thanks to the members of the Singular Optics Laboratory at IIT Delhi for their assistance. PS acknowledges the financial support provided by the Science and Engineering Research Board (SERB), India, under Grant No. CRG/2022/001267.
\end{acknowledgments}
\appendix
\bibliography{su2}
\end{document}